# Experimental Demonstration of Autonomous Class Aware Optimization via Digital Twin Aided Optical Network Agent

Marc Lawson(1,2), Ambashri Purkayastha(2,3), Camille Delezoide(2), Andrea Sgambelluri(1), Nicola Sambo(1), Patricia Layec(2)

(1)Scuola Superiore Sant'Anna, Pisa, Italy, marc.lawson@santannapisa.it
(2) Nokia Bell Labs, Massy, France
(3) LTCI, Télécom Paris, Institut Polytechnique de Paris, France

**Abstract** *We demonstrate on a network testbed a Digital Twin-aided multi-agent LLM for autonomous class-aware optical network management. The system detects soft failures and autonomously adapts modulation format per service class requirements, ensuring continuous, marginless channel quality with minimal human intervention.* 

## Introduction

Optical networks have accelerated the adoption of Agentic LLM frameworks capable of supporting closed-loop control, intent-driven operation, and real-time decision-making [1]. These functionalities are fundamental in the context of marginless operations [2], for which a network is pushed more and more towards its limits, requiring proper performance observation and reactions in case of problems. Recent demonstrations highlight the feasibility of LLM-powered agents for autonomous driving optical networks, where coordinated agents manage wavelength add/drop operations, detect soft and hard failures, and perform power optimization across the network lifecycle [3]. Complementary work has shown that multi-agent LLMs can enhance digital-twin-based workflow automation by integrating intent parsing, telemetry analysis, auto-simulation, and closed-loop optimisation, enabling QoT-aware provisioning and self-improving reasoning in operational environments [4]. These advances illustrate the growing role of multi-agent LLM architectures in decomposing complex optical-network tasks into specialized roles — planning, telemetry interpretation, configuration execution, and validation — improving reliability and reducing operator intervention.

However, a fully autonomous multi-agent LLM framework with SDN control for service optimization is yet to be presented.

This work presents an LLM-based framework for autonomous soft failure detection and correction — an essential capability for marginless operation — endowed with SDN control. We demonstrate a fine-tuned multi-agent LLM for reasoning and decision-making, supported by a Digital Twin (DT), that reliably executes QoT-driven decisions for marginless networking. We experimentally validate that the system autonomously detects soft failures before they escalate and leverages DT-assisted BER predictions to maintain fully automated operation of marginless connections over a 630 km path under emulated amplifier and fiber ageing.

## Multi-Agent LLM Architecture

The control architecture — shown in Fig. 1 — instantiates a plan–verify–execute paradigm with numbered workflow in the figure tracing the full decision loop. The workflow is initiated at the human–machine interface, where natural language input (1) (e.g., "in case of soft failure, run modulation-format adaptation") is received by the Supervisor Agent — a fine-tuned LLaMA3.1-8B responsible for intent parsing and task delegation. The Supervisor forwards the interpreted task (2) to the Toolbox, a centralized orchestration bus that simultaneously ingests real-time telemetry (3) from a NETCONF/YANG-polled Telemetry Database, ensuring all downstream actions are grounded in current network state. The Toolbox dispatches tasks (4) to two specialized agents: the Config Agent translates high-level intent into vendor-neutral NETCONF remote procedure calls (RPCs) for amplifier gain, variable optical attenuator (VOA), and wave-length selective switch (WSS) settings; the QoT Agent performs quality-of-transmission estimation and parameter optimization based on detected impairments.

Both are fine-tuned LLaMA3.1-8B models that query (5) a shared Knowledge Base — containing standards, vendor specifications, and physical-layer constraints — via Retrieval-Augmented

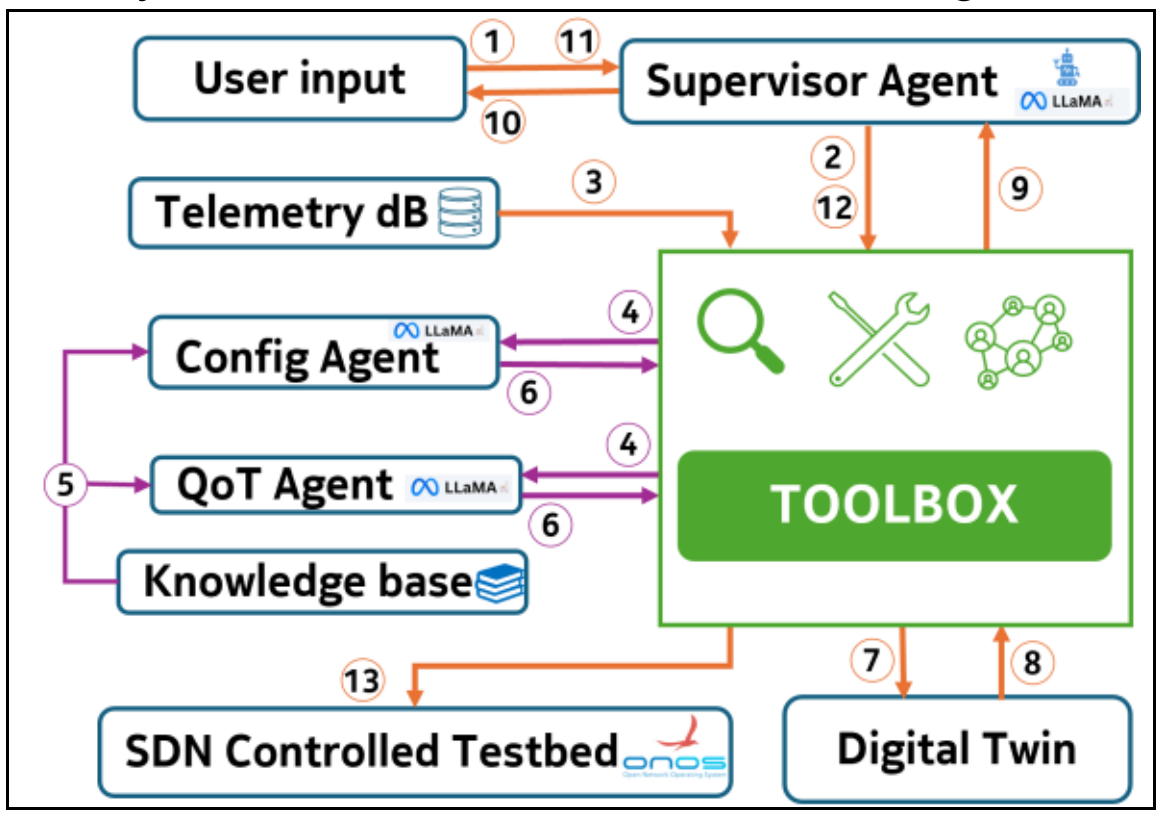

Fig. 1: Proposed multi-agent LLM architecture

| (a) Model | | MMLU | ARC | Perplexity |
|---|---|---|---|---|
| LLaMA 2-7B | Base | 0.449 | 0.437 | 9.957 |
| | FT | 0.450 | 0.442 | 8.808 |
| | Δ (%) | **+0.25** | **+1.16** | **+11.55** |
| LLaMA 3-8B | Base | 0.639 | 0.518 | 8.640 |
| | FT | 0.641 | 0.543 | 5.469 |
| | Δ(%) | **+0.28** | **+4.56** | **+36.70** |
| Mistral 7B | Base | 0.601 | 0.507 | 6.088 |
| | FT | 0.598 | 0.529 | 4.187 |
| | Δ(%) | **+0.49** | **+4.19** | **+31.22** |

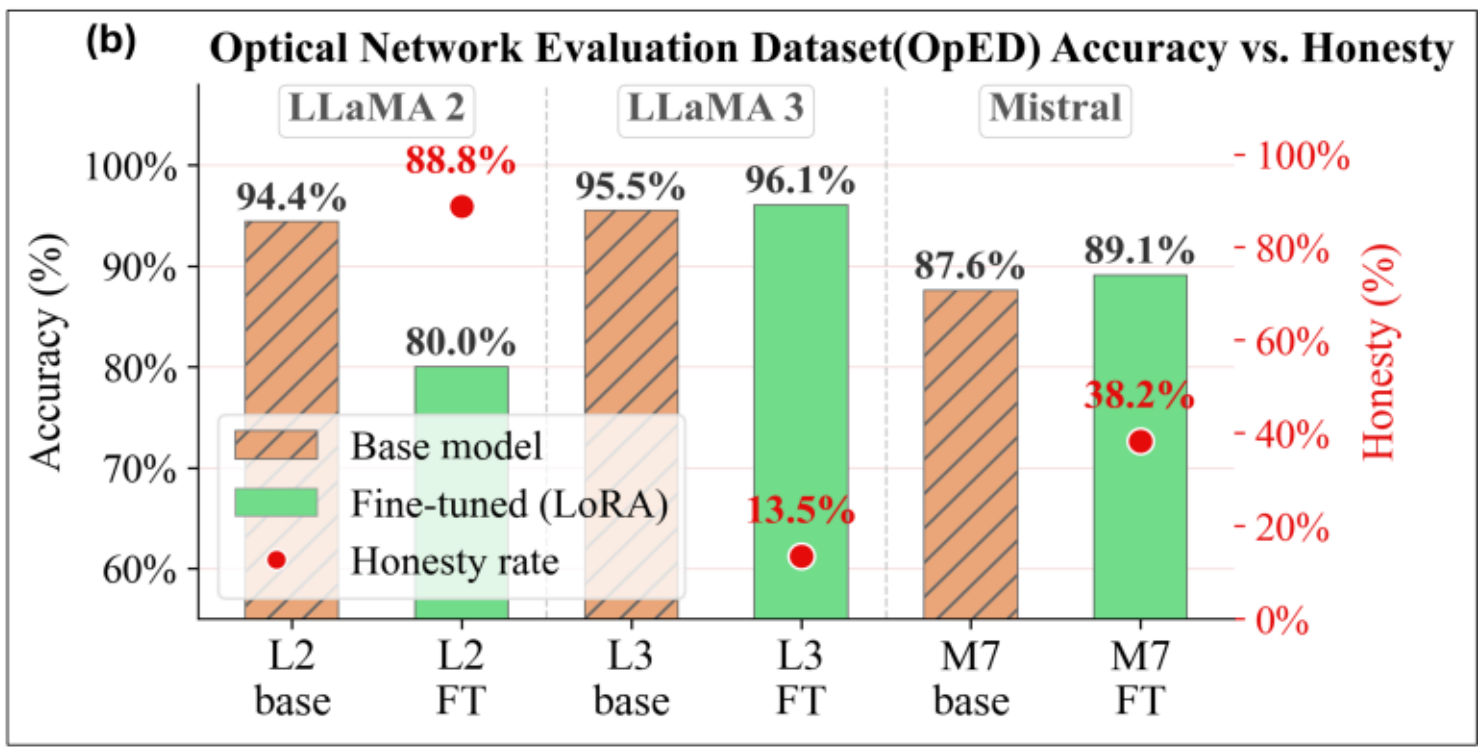

**Fig. 2:** (a) Standard benchmarking evaluation - base and fine-tuned (FT) LLMs; (b) Evaluation of base and fine-tuned LLMs

Generation (RAG), an inference-time mechanism that conditions model outputs on retrieved domain context to suppress hallucination [5]. Agent responses are returned (6) to the Toolbox.

Before any configuration is applied, the Toolbox forwards candidate actions (7) to the Digital Twin, which validates them against physical-layer constraints — gain limits, margin requirements, and forward error correction (FEC) thresholds — and returns the verification result (8). The outcome is reported back (9) to the Supervisor Agent, which communicates the decision (10) to the user; if further iteration is required, the loop re-enters (11, 12) until a valid action is confirmed. Only verified configurations are dispatched (13) to the SDN controller (an ONOS-based controller is exploited in this work)—ensuring physically inconsistent or overconfident LLM outputs are intercepted before reaching live hardware.

Three models — LLaMA2-7B [6], LLaMA3.1-8B [7], and Mistral-7B [8] — were fine-tuned via QLoRA [9] on testbed telemetry and ITU-T standards documentation, then evaluated on LLM benchmarks: Massive Multitask Language Understanding (MMLU, i.e., a set of multiple-choice questions) [10], AI2 Reasoning challenge (ARC, i.e. dataset designed to evaluate the reasoning capabilities) [11], and perplexity (a measure of how accurately a model predicts the next token in a sequence) [12] – and secondly on *OpED* — the Optical Network Evaluation Dataset, a domain-specific benchmark developed in this work covering optical hardware, DSP, and SDN/NFV automation, with a honesty penalty for hallucinated responses. LLaMA3.1-8B fine-tuned achieved the strongest results across both evaluation axes: the largest perplexity reduction at 36.70% on standard benchmarks (Fig. 2a), and the highest OpED accuracy at 96.1% (Fig. 2b). Its low honesty rate of 13.5% reflects a small number of abstained questions, the majority of which were answered correctly — unlike base models, which attempted all questions regardless of confidence and yielded no honesty signal. Given its overall superiority, LLaMA3.1-8B fine-tuned was selected for deployment across all agent roles.

## Experimental demonstration and results

The physical layer testbed in Fig. 3, consists of 8 commercial-grade reconfigurable add/drop multiplexer (ROADM) nodes, connected in a mesh topology. We transmit two channels from two different generations of commercial coherent transceivers from the same vendor, that are operating under two different service class agreements (SLA): *gold* class (high priority) and *bronze* class (low priority). Gold SLA is to maintain constant net rate with a high margin, ensuring robust service delivery. Conversely, Bronze SLA allows the channel to operate at zero-margin, with the caveat that the net rate can be sacrificed under strenuous service conditions.

We have continuous monitoring of the pre-FEC bit error rate (BER), along with the OSNR at each node, total spectrum and channel powers at the input and output of each node, amplifier gain, VOA and wavelength selective switch (WSS) attenuations from the ROADMs. The telemetry data is stored in the InfluxDB and is visualized on the Grafana interface. As described in Fig.3, the physical layer data plane is controlled via the interfaced with a ONOS-based SDN controller.

The multi-Agent architecture employs Python functions to handle analytics while the LLM provides higher level reasoning.

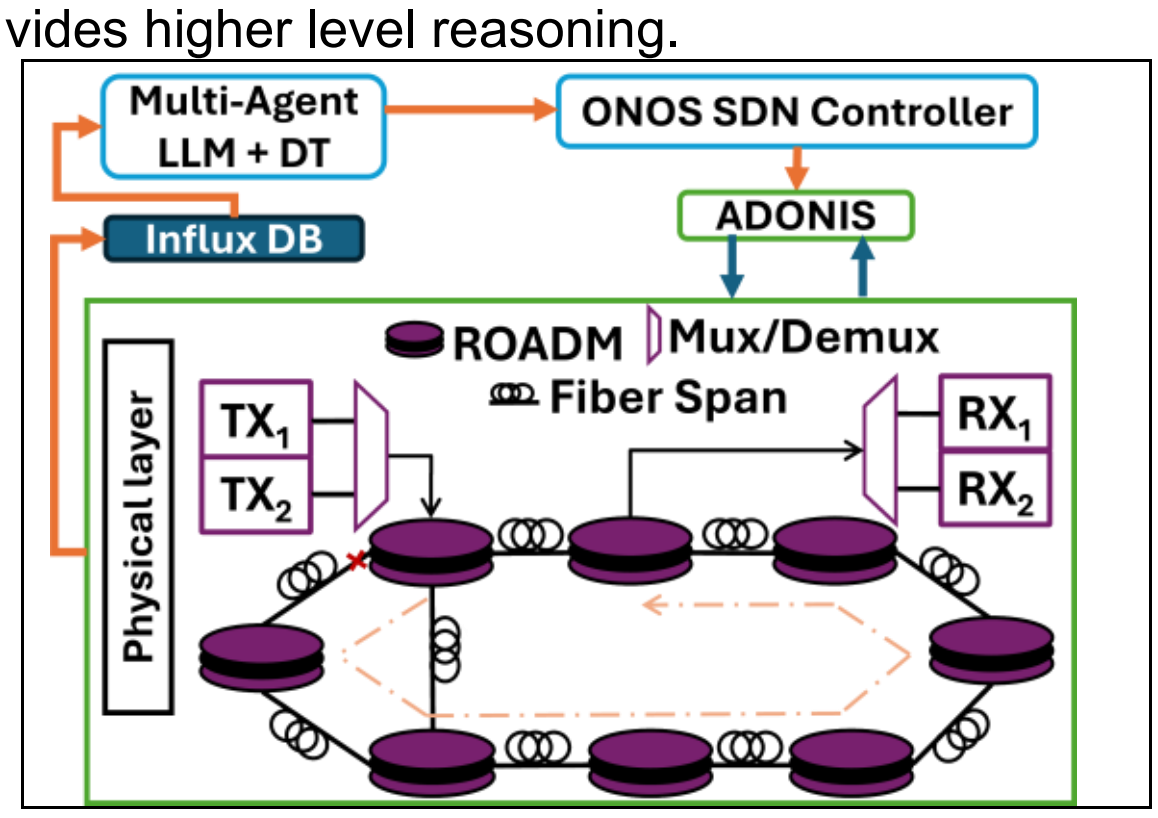

Fig. 3: Experimental testbed topology

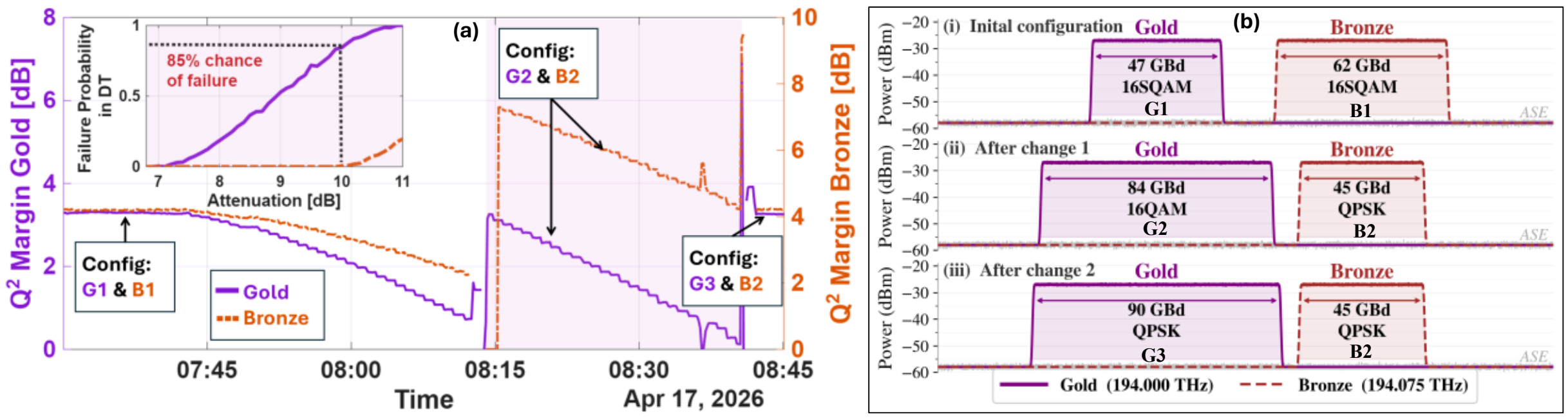

Fig. 4: (a) Timeseries of mode transitions for gold and bronze service after verification of performance within the DT; (b) Optical spectrum, before and after service-based mode adaptation

The QoT Agent monitors pre-FEC-BER against a soft failure threshold $Th_{SF} = \langle BER \rangle + 4\sigma$ (from a 9-sample window and standard deviation $\sigma$ obtained from live measurements) triggering adaptation when BER reaches about 20% of the FEC margin. A failure detection algorithm selects a lower-rate candidate mode and evaluates feasibility via: FEC headroom ($\log_{10}$ FEC_limit/BER), BER trend, and BER prediction using calibrated fiber attenuation coefficients $\alpha_k$ from [2]. This assessment plus cross-channel context is passed to the LLM for an *APPROVE / DEFER / REJECT* decision. If approved, the Config Agent applies the new mode via ONOS SDN Controller.

To assess the framework's robustness, we emulate fiber ageing by increasing VOA attenuation at ROADM-3 from 0 dB to 12 dB in steps of 0.2 dB, holding each level for 5–10 minutes to ensure steady-state conditions. The attenuation is initialized at 0 dB so that the LLM agents capture the baseline BER distribution and compute $Th_{SF}$. As the loss increases, the Q-factor calculated from the pre-FEC BER from [13] drops and the available FEC margin defined as $Q^2_{margin} = Q^2_{measured} - Q^2_{FEC}$ (in dB) narrows, as shown in Fig. 4(a) for both services: Gold and Bronze.

The QoT Agent continuously streams the measured BER to its LLM-based decision layer. When the BER approaches approximately 80 % of the allowable FEC margin, the LLM triggers the operational mode-adaptation algorithm.

By reducing the symbol rate of the Bronze service, spectrum is freed for the Gold service to maintain the same net rate, with the inherent assumption that both channels co-propagate through the same nodes until reception. The Configuration Agent verifies the proposed configuration on the DT, which emulates statistical network performance under the proposed trend failure. The DT then returns the expected $Q^2$ with failure probability when the $Q^2$ is above the FEC margin, built from past measurements following the method in [9]. Based on this response, the Configuration Agent selects the most optimum combination to satisfy both SLAs and deploys it on the physical network via ONOS.

The resulting transitions are shown in Fig. 4(a): with increasing attenuation, the Gold service initially steps from G1 to G2 by forcing the Bronze service to downgrade from B1 to B2. With further attenuation, the failure probability for Gold service calculated by the DT reaches 85% (see inset), which forces another change from G2 to G3, while B2 remains unchanged as shown in Fig 4(b).

The illustration in Fig.5(a), ONOS log records the event sequence (SOFT_FAILURE, CONFIG_INTENT, and CONFIG_APPLIED) during testbed configuration. Fig. 5(b) details the LLM assessment, including decision rationale, confidence level, and the operator alert.

20:33:30.161 INFO [core] [ADONIS-AGENT] event=SOFT_FAILURE, channel=OCH-1-26-L1, port=1/26/L1, device=Gold_class_-_TRX-2, ber=2.126e-02,ber_pct=7.870e+01, zone=SAFE, action=APPROVE, port=1-26-L1 || 20:34:58.637 INFO [core] [ADONIS-AGENT]event=CONFIG_INTENT, channel=OCH-1-16-L1, port=1/16/L1, device=Bronze_class_TRX-1, mode_id=B2, freq_ghz=1.941e+05, format=QPSK, port=1-16-L1 || 20:35:24.750 INFO [core] [ADONIS-AGENT] event=CONFIG_APPLIED, channel=OCH-1-16-L1, port=1/16/L1,device=Bronze_class_TRX-1, mode_id=B2, freq_after=194075.000, mode_after=B2, status=ok (a)

LLM ASSESSMENT - Soft failure: True | Emergency: False | Action: APPROVE | Direction: DOWNGRADE | Mode selected: G3 | Feasibility: SAFE | Confidence: 50% | Reason: BER 2.202e-02 < FEC margin || [ALERT TO OPERATOR]:Channel OCH-1-26-L1 was modified from 16SQAM 300 Gbps 84 Gbaud to QPSK 300 Gbps 92 Gbaud. (b)

Fig. 5: (a) ONOS SDN logs for configs; (b) LLM assessment with decision rationale and operator alert.

## Conclusion

This paper fine-tunes and evaluates LLM-based agents integrated into a multi-agent framework for autonomous soft-failure detection and correction in optical networks. A DT serves as a verification layer to mitigate LLM hallucinations. We experimentally demonstrated consistent service optimization, by performing timely mode-adaptation capable of autonomously maintaining service class requirements under progressive degradation, without requiring manual operation.